\documentstyle[amssymb,aps]{revtex}

\begin{document}
\draft
\title{Doubly--dressed atomic wave packets}
\author{G. Yu. Kryuchkyan}
\address{Yerevan State University, Alex Manookian 1, 375049, Yerevan, Armenia}
\maketitle

\begin{abstract}
The problems of cavity atom optics in the presence of an external strong
coherent field are formulated as the problems of potential scattering of
doubly-dressed atomic wave packets. Two types of potentials produced by
various multiphoton Raman processes in a high-finesse cavity are examined.
As an application the deflection of dressed atomic wave packets by a cavity
mode is investigated. New momentum distribution of the atoms is derived that
depends from the parameters of coherent field as well as photon states in
the cavity.
\end{abstract}

\section{Introduction}

Recent technological achievements in quantum optics in that number
improvements in cavity design initiated the investigations of a single atom
dynamics via its interaction with quantized cavity field. These studies have
opened a new chapter of atom optics where we treat not only the states of
the atom and its motion quantum mechanically, but have also quantized the
cavity field [1-3]. Among the models of atom-cavity mode interactions most
fascinating ones is related to potential problems in quantum theory. As was
shown in context of micromazers [4-8], the resonant interaction between an
incident atom and a cavity mode can be formulated as the elementary problem
of a particle incident upon potential, if irreversible processes which give
rise to damping, such as atomic spontaneous emission and decay of the field
mode, are negligible. The cavity field in this approach acts as a potential
barrier or as a potential well for the dressed-states of the atom-cavity
field system.

In this paper we extend borders of the above noted analogy between atomic
dynamics and scattering problem considering interaction of an atom with both
a quantized cavity mode and a classical strong laser field. This situation
may be realized, in particular, in the following scheme shown on Fig.1,
where the atomic probe propagates through a cavity and the laser field
travels orthogonally with respect to the cavity mode. In such schemes the
moving atom interacts at the same time with a near-resonant, strong coherent
field and a cavity mode. The atom is slow enough that adiabatic switching on
the coherent field-atom interaction is realized and the position dependent
dressed-states of the atom-coherent field system are formed [9]. As we show
bellow, the srong coupling of a single atom to a single cavity mode induces
transitions between these position dependent dressed states which can be
explained through the cavity field dressing of the strong field-atom dressed
states. Such approach allows us to formulate the problem of atoms coupled
with quantized cavity mode in the presence of strong laser field as the
problem of potential scattering of doubly-dressed atomic wave packet.

The application of doubly-dressed states to the problem of atomic deflection
is the other purpose of this paper. Atomic deflection during its passage
through the cavity has been studied extensively. A detail review on this
field can be found in Ref. [1]. In particular, it was shown that the atomic
deflection pattern contains information about the field state in the cavity
and can be used for the probing and reconstruction of quantum states
[10-13], and for subject of an atomic interferometry [14]. Here, we study
the influence of the external coherent field on atomic deflection pattern.

\section{Potentials for doubly-dressed states}

We consider dynamics of an incident two-level atom with ground state $\mid
g\rangle $ and excited state $\mid e\rangle $ moving along z-direction and
passes a light field. This field consists of a standing quantized cavity
mode in an arbitrary state and a running coherent field. We assume the time
of interaction between the atom and the fields to be very short compared to
the cavity lifetime $1/k$ and the inverse spontaneous decay rate $\gamma $ ,
that is $t\ll 1/k$, $1/\gamma $. The effective Hamiltonian of the atom
placed at the point $r(x,y,z)$ in a cavity in the rotating-wave
approximations reads 
\begin{equation}
H=H_{0}+H_{R}+H_{ext}+H_{int\text{ }}.  \label{1}
\end{equation}
$H_{0}$ and $H_{R}$ are the free Hamiltonians for the atom and the quantized
cavity mode at frequency $\omega _{c}$

\begin{equation}
H_{0}=\frac{\widehat{p}^{2}}{2m}+\omega _{g}\mid g\rangle \langle g\mid
+\omega _{e}\mid e\rangle \langle e\mid ,
\end{equation}

\begin{equation}
H_{R}=\hbar \omega _{c}a^{+}a,
\end{equation}
where $\widehat{p}$ denotes the center-of-mass motion momentum operator of
the atom with mass $m,$ and $\omega _{e}$ and $\omega _{g}$ are the energies
of the excited and ground electronic states of the atom, respectively. The
Hamiltonian 
\begin{equation}
H_{ext}=\hbar \lambda u_{L}(r)(\sigma _{+}e^{-i(\omega _{L}t+\varphi
_{L})}+h.c.)  \label{2}
\end{equation}
describes the interaction between the atom and the classical traveling field
of frequency $\omega _{L}$ and phase $\varphi _{L}$, where $\lambda $ is the
coupling constant, $\sigma _{+}$ is the atomic flip operator and $u_{L}(r)$
is the position- dependent part of the external field amplitude. The
interaction of the atom with a field mode reads 
\begin{equation}
H_{int}=\hbar gu(r)(\sigma _{+}ae^{-i\varphi _{c}}+h.c.),  \label{3}
\end{equation}
where $g$ is the coupling constant and the field mode is determined by the
spatial mode function $u(r)$, by a phase $\varphi $ and the annihilation and
creation operators $a$, $a^{+}$.

We use an interaction-picture version in which the equation for vector state
of the total system is 
\begin{equation}
i\hbar \frac{\partial }{\partial t}\mid \Psi (t)\rangle =\left( \frac{%
\widehat{p}^{2}}{2m}+\hbar gu(r)\left( \sigma _{+}(t)ae^{-i\omega
_{c}t-\varphi _{c}}+\sigma _{-}(t)a^{+}e^{i\omega _{c}t+\varphi _{c}}\right)
\right) \mid \Psi (t)\rangle \text{,}  \label{4}
\end{equation}

where 
\begin{equation}
\sigma _{\pm }(t)=\exp [\frac{i}{\hbar }\int_{-\infty
}^{t}(H_{0}+H_{ext}(t^{\prime }))dt^{\prime }]\sigma _{\pm }\exp [-\frac{i}{%
\hbar }\int_{-\infty }^{t}(H_{0}+H_{ext}(t^{\prime }))dt^{\prime }]
\label{8}
\end{equation}
are the atomic states flip operators in this representation. It is useful to
describe the atom-coherent field subsystem in terms of position-dependent
Floquet states which for two-level atom at point $r$ and in the
rotating-wave approximation are \ \ \ \ \ \ \ \ \ \ \ \ \ \ \ \ \ \ \ \ 
\begin{eqnarray}
&\mid &\Psi _{1}(r,t)\rangle =e^{-i\omega _{1}(r)t}\left( a\mid g\rangle
+b\mid e\rangle e^{-i(\omega _{L}t+\varphi _{L})}\right) ,  \nonumber \\
&\mid &\Psi _{2}(r,t)\rangle =e^{-i\omega _{2}(r)t}\left( -b\mid g\rangle
e^{i(\omega _{L}t+\varphi _{L})}+a\mid e\rangle \right) .
\end{eqnarray}
The quasienergies $\hbar \omega _{1}(r)$ and $\hbar \omega _{2}(r)$ are
equal to 
\begin{eqnarray}
\omega _{1}(r) &=&\omega _{g}+\frac{1}{2}\left( \delta -\Omega (r)\right) , 
\nonumber \\
\omega _{2}(r) &=&\omega _{e}-\frac{1}{2}\left( \delta -\Omega (r)\right) ,
\end{eqnarray}
where the position-dependent Rabi frequency is $\Omega (r)=\sqrt{\delta
^{2}+4\lambda ^{2}u_{L}(r)^{2}}$, and the coefficients $a,b$ are defined by $%
a=\sqrt{\frac{1}{2}(1+\delta /\Omega )},\,\,\,\,\,\,b=$ $\sqrt{\frac{1}{2}%
(1-\delta /\Omega )},$ where the detuning $\delta =\omega _{e}-\omega
_{g}-\omega _{L}\ll \omega _{e}-\omega _{g}$. The important point is that
the states (8) will vary with the position $r$. They describe the
atom-coherent field subsystem in the absence of spontaneous emission and for
an atomic velocity sufficiently small that nonadiabatic transitions between
these states can be ignored. In order to gain insight into the dynamics of
the system it is convenient to operate with the states $\mid \Phi
_{1}(r)\rangle =\mid \Psi _{1}(r,0)\rangle $ and $\mid \Phi _{2}(r)\rangle
=\mid \Psi _{2}(r,0)\rangle $, 
\begin{eqnarray}
&\mid &\Phi _{1}(r)\rangle =a(r)\mid g\rangle +b(r)e^{-i\varphi _{L}}\mid
e\rangle , \\
&\mid &\Phi _{2}(r)\rangle =-e^{i\varphi _{L}}b(r)\mid g\rangle +a(r)\mid
e\rangle ,
\end{eqnarray}
which are similar to the position-dependent dressed states introduced for a
moving atom in Ref.[9]. Then, the state vector $\mid \Psi (t)\rangle $ of
the total \ \ system can be expressed in the basis of such type of
dressed-states and photon number states $\mid \Phi _{i}(r)\rangle \mid
n\rangle $ and in the position representation is given as 
\begin{equation}
\mid \Psi (t)\rangle =\sum_{i=1,2}\sum_{n=0}^{\infty }\int
d^{3}rf_{i}^{n}(r,t)\mid \Phi _{i}(r)\rangle \mid n\rangle \mid r\rangle 
\text{.}  \label{12}
\end{equation}

Substituting the quantum state (12) in Eq.(6) we find the evolution of the
probability amplitudes in the following form 
\begin{eqnarray}
i\hbar \frac{\partial }{\partial t}f_{j}^{n}(r,t) &=&\frac{\widehat{p}^{2}}{%
2m}f_{j}^{n}(r,t)+\hbar gu(r)\sum_{i}f_{i}^{n-1}(r,t)\sqrt{n}e^{i(\omega
_{c}t+\varphi _{c})}A_{ji}(r,t)+ \\
&&\hbar gu(r)\sum_{i}f_{i}^{n+1}(r,t)\sqrt{n+1}e^{-i(\omega _{c}t+\varphi
_{c})}A_{ij}^{\ast }(r,t)+(F_{j}^{n})_{NA},  \nonumber
\end{eqnarray}
where the coefficients $A_{ji}(r,t)=\langle \Phi _{j}(r)\mid \sigma
_{-}(t)\mid \Phi _{i}(r)\rangle $ are obtained as 
\begin{eqnarray}
A_{11} &=&-A_{22}=ab\exp [-i(\omega _{L}t+\varphi _{L})], \\
A_{12} &=&a^{2}\exp [-i(\omega _{L}+\Omega )t],A_{21}=-b^{2}\exp [-i(\omega
_{L}+\Omega )t-2i\varphi _{L}].  \nonumber
\end{eqnarray}
\ The term $(F_{j}^{n})_{NA}$ represents the contribution of nonadiabatic $%
(NA)$ part induced by the transitions between dressed states of a moving
atom in the absence of photon emission. They appear due to both the spatial
and the temporal variations of the dressed states because the point r varies
with time. Using that $\partial \mid \Phi _{i}(r(t))\rangle /\partial t=%
\overrightarrow{\upsilon }\nabla \Phi _{i}(r(t))\rangle $, where $%
\overrightarrow{\upsilon }$ is the atomic velocity, the\ $NA$ coupling can
be written as 
\begin{equation}
(F_{j}^{n})_{NA}=\sum_{i}f_{i}^{n}(r,t)\langle \Phi _{j}\mid \frac{\widehat{p%
}{{}^{2}}}{2m}\mid \Phi _{i}\rangle +\frac{1}{2m}\sum \langle \Phi _{j}\mid 
\widehat{p}\mid \Phi _{i}\rangle \,\widehat{p}f_{i}^{n}(r,t)-i\hbar
\sum_{i}f_{i}^{n}(r,t)\overrightarrow{\upsilon }\langle \Phi _{j}\nabla \Phi
_{i}\rangle \text{.}  \label{16}
\end{equation}

We consider below an enough slow moving atom for which nonadiabatic effects
are negligible. Let us give an order of magnitude the longitudinal velocity $%
\upsilon _{z}$ of atomic wave packet for which the $NA$ coupling is
negligible compared with the spontaneous transitions. Following [9] for
analyzing of nonadiabatic transitions and assuming a plane wave coherent
field with the Gaussian transverse spatial distribution $%
u_{L}(r)=e^{ik_{y}y}(\pi \Delta z^{2})^{1/4}\exp (-z^{2}/2\Delta z^{2})$ we
arrive at 
\begin{equation}
\upsilon _{z}\ll \gamma ^{1/3}\frac{\Delta z\Omega ^{2}(0)}{(\delta \lambda
u(0))^{2/3}},  \label{18}
\end{equation}
where $\Omega (0)$ is a maximal value of the position-dependent Rabi
frequency at the cavity centre $r=0.$ In this limit the system adiabatically
follows the dressed states and the last term of the equation (13) can be
neglected. Therefore, \ the resulting equation involves only the radiative
transitions.

The spectral lines of driving atom have well-known three-peak structure at
the coherent field frequency $\omega _{L}$ and two sidebands $\omega _{L}\pm
\Omega $ which are symmetrically displaced about the central peak by the
position-dependent Rabi frequency $\Omega $. This circumstance makes it
possibly to consider several atom-cavity mode couplings. There are two
situations deserve attention occurring if the cavity frequency is resonant
with one of two spectral lines of driving atom $(\omega _{c}=\omega
_{L}-\Omega $ or $\omega _{c}=\omega _{L}+\Omega )$.

\subsection{ ''Three photon'' coupling}

Let cavity is tuned at the red spectral sideband of the driven atom. We
assume the resonance condition $\omega _{c}=\omega _{L}-\Omega (0)$ to be
satisfied for the maximum value $\Omega (0)$ of the position-dependent Rabi
frequency and make the resonant approximation in Eq.(13) keeping the slowly
oscillating terms $e^{i\Delta (r)t}$, where $\Delta (r)=\Omega (r)-\Omega
(0) $. This procedure gives the system of coupled equations for the
amplitudes $f_{1}^{n-1}$ and $f_{2}^{n}$. We can decuple these two equations
by introducing the linear combinations of the amplitudes 
\begin{equation}
\phi _{n}^{(\pm )}(r,t)=f_{1}^{n-1}(r,t)\pm e^{i(2\varphi _{L}-\varphi
_{c}-\Delta t)}f_{2}^{n}(r,t),
\end{equation}
(for $n\geq 1$). As it can be shown each of these components obeys the
time-dependent Schr\"{o}dinger equation 
\begin{equation}
i\hbar \frac{\partial }{\partial t}\phi _{n}^{(\pm )}(r,t)=\left( \frac{%
\widehat{p}{{}^{2}}}{2m}+U_{n}^{(\pm )}(r)\right) \phi _{n}^{(\pm )}(r,t)
\label{20}
\end{equation}
with the potential 
\begin{equation}
U_{n}^{(\pm )}(r)=\mp \frac{\hbar }{2}gu(r)(1-\delta /\Omega (r))\sqrt{n},
\label{21}
\end{equation}
while for the zero-photon number the equations (13) give

\begin{equation}
i\hbar \frac{\partial }{\partial t}f_{2}^{0}=\frac{\widehat{p}{{}^{2}}}{2m}%
f_{2}^{0}.
\end{equation}

We have thus reduced the problem of atomic dynamics in a high-finesse cavity
in the presence of external coherent field to the elementary scattering
process. Rewriting the quantum state (12) in terms of the amplitudes (17)
and projecting its onto the position eigenstate $\mid r\rangle $ we obtain 
\begin{equation}
\langle r\mid \Psi (t)\rangle =f_{2}^{0}(r,t)\mid \Phi _{2}\rangle \mid
0\rangle +\frac{1}{\sqrt{2}}\sum_{n\geq 0}\left( \phi _{n+1}^{(+)}(r,t)\mid
N_{n}^{(+)}\rangle +\phi _{n+1}^{(-)}(r,t)\mid N_{n}^{(-)}\rangle \right) ,
\end{equation}
where 
\begin{equation}
\mid N_{n}^{(\pm )}\rangle =\frac{1}{\sqrt{2}}\left( \mid \Phi _{1}\rangle
\mid n\rangle \pm e^{i(\Delta t+\varphi _{c}-2\varphi _{L})}\mid \Phi
_{2}\rangle \mid n+1\rangle \right) ,
\end{equation}
for $n\geqslant 0.$ So, the quantum state of the system is obtained in terms
of the orthogonal basis of doubly--dressed states $\mid N_{n}^{(\pm
)}\rangle $ which are formed as the linear combinations of the coherent
field--atom dressed states multiplied on the occupation numbers of cavity
mode. Note, that the emission and absorption of photons at the frequency $%
\omega _{c}=\omega _{L}-\Omega (0)$ are stipulated by well known Raman
scattering processes. For large detuning $\delta ^{2}\gg \lambda
^{2}u_{L}^{2}(0)$ and in the lowest order of perturbation theory on coherent
field-atom interaction these processes are shown in Fig.2. The atom radiates 
$\omega _{L}-\Omega $ photon in the transition from the ground state $\mid
g\rangle $ to the excited state $\mid e\rangle $ in which two photons of the
coherent field are absorbed (Fig.2(a)). Absorption of $\omega _{L}-\Omega $
photon from the field mode takes place in the transition $\mid e\rangle
\rightarrow \mid g\rangle $ with the radiation of two photons at frequency $%
\omega _{L}$ (Fig.2(b)). Such Raman processes form the fluorescence spectrum
of strongly driven two-level atoms into the mode of an optical cavity as has
been experimentally demonstrated in [15]. For moving atom in a cavity the
potential (19), which for the large detuning reads as 
\begin{equation}
U_{n}^{(\pm )}(r)=\mp \frac{\hbar g\lambda ^{2}}{\delta ^{2}}u(r)u_{L}(r)^{2}%
\sqrt{n},  \label{25}
\end{equation}
is induced by the three-photon Raman processes.

\subsection{Resonance $\protect\omega _{c}=\protect\omega _{L}+\Omega $
condition}

In that resonant case Eqs.(13) are transformed to the coupled equations for
the amplitudes $f_{1}^{n}$ and $f_{2}^{n-1}$. We decuple these equations by
introducing the following combinations 
\begin{equation}
\theta _{n}^{(\pm )}(r,t)=f_{1}^{n}(r,t)+e^{i(\varphi _{c}-\Delta
t)}f_{2}^{n-1}(r,t),\,\,  \label{26}
\end{equation}
(for $n\geq 1$). Then the equations are simplified as 
\begin{equation}
i\hbar \frac{\partial }{\partial t}\theta _{n}^{(\pm )}(r,t)=\left( \frac{%
\widehat{p}{{}^{2}}}{2m}+V_{n}^{\left( \pm \right) }(r)\right) \theta
_{n}^{(\pm )}(r,t),  \label{27}
\end{equation}
where the potential is equal to 
\begin{equation}
V_{n}^{\left( \pm \right) }(r)=\pm \frac{\hbar }{2}gu(r)(1+\delta /\Omega
(r))\sqrt{n}.  \label{28}
\end{equation}
For $n=0$ the equation is written as

\begin{equation}
i\hbar \frac{\partial }{\partial t}f_{1}^{0}=\frac{\widehat{p}{{}^{2}}}{2m}%
f_{1}^{0}.
\end{equation}

In a similar way we transform the quantum state (12) in terms of the
amplitudes (24) and the doubly-dressed states. The result is calculated as 
\begin{equation}
\langle r\mid \Psi (t)\rangle =f_{1}^{0}(r,t)\mid \Phi _{1}\rangle \mid
0\rangle +\frac{1}{\sqrt{2}}\sum_{n\geq 0}\left( \theta
_{n+1}^{(+)}(r,t)\mid R_{n}^{(+)}\rangle +\theta _{n+1}^{(-)}(r,t)\mid
R_{n}^{(-)}\rangle \right) ,  \label{29}
\end{equation}
through the following doubly--dressed states 
\begin{equation}
\mid R_{n}^{(\pm )}\rangle =\frac{1}{\sqrt{2}}(\mid \Phi _{1}\rangle \mid
n+1\rangle \pm e^{i(\Delta t-\varphi _{c})}\mid \Phi _{2}\rangle \mid
n\rangle )  \label{30}
\end{equation}
for $n\geq 0.$

So, the quantum states (21), (22) and (28), (29) show a strong entanglement
between the atom and the light field involved the coherent component and the
single cavity mode. The cavity with fixed number of photon acts as the
potential barrier or as the potential well for the doubly-dressed states of
atom. It should be mentioned that the potentials (19) and (26) show
characteristic dependence on both atom-light field coupling constants and
the detuning of the coherent field.

The approach followed above is well adopted to the strong-coupling regime,
where the coupling constants $\lambda ,g$ are greater than the decay rates
of the atomic dipole $\gamma $ and the cavity field $k$ ($\lambda \gg \max
[\gamma ,k],g\gg \max [\gamma ,k])$. In this regime and for the time
intervals $\max [g^{-1},\lambda ^{-1}]\ll t\ll \max [\gamma ^{-1},k^{-1}]$
the atomic relaxation are negligible during the atom transit time across the
cavity. Another peculiarity of this regime is that three spectral lines of
the atomic radiation into cavity modes at frequencies $\omega _{L},\omega
_{L}-\Omega ,\omega _{L}+\Omega $ are well resolved. It makes it possible to
chose the cavity resonance frequency equal to one of the sidebands. The
equations (18 ), (25 ) are valid for the velocities satisfying the adiabatic
approximation. It is easy to see from (16) that for a resonant coherent wave 
$(\delta =0)$ the adiabatic approximation holds for any velocity.

\section{Deflection of the dressed atom}

Consider now the deflection of dressed atomic wave packet by a quantized
electromagnetic field in the Raman-Nath regime. In the scheme shown on Fig.1
we assume a sinusoidal spatial mode function $u(x,y,z)=\sin (kx)$ for the
quantized standing wave and treat the laser field as a wave packet $%
u_{L}(x,y,z)=u_{L}(x,z)\exp (ik_{L}y)$ with the wave vector $k_{L}$ and with
the sufficiently broadband transverse profile $u_{L}(x,z)$ so that the
adiabatic approximation holds.

The atomic motion along the z-axis is treated classically that implies that
the initial kinetic energy $p_{z}^{2}/2m$ is much larger than the change of
the longitudinal momentum due to the interaction. The Raman-Nath
approximation means that during the interaction time we neglect the transfer
of kinetic energy in the x-and the y-directions to the atom. These
approximations make it possible to reduce, the equations (18), (25) to the
one-dimensional equations, where the coordinate $z=\upsilon _{z}t$ is
proportional to the interaction time.

Suppose now a two-level atom in the ground state $\mid g\rangle $ and with
transverse centre-of-mass wave function $f(x)$ enters a cavity with a
single-mode standing light in the quantum state $\sum c_{n}\mid n\rangle .$
The initial state vector reads 
\begin{equation}
\mid \Psi (t=0)\rangle =\sum_{n\geq 0}\int dxf(x)c_{n}\mid x\rangle \mid
g\rangle \mid n\rangle .  \label{31}
\end{equation}
It is known that the spatial periodicity of the standing wave leads to
discrete atomic momenta with a spacing of $\hbar k$. Our aim is to find the
corresponding momentum distribution of the atom due to its interactions with
both the coherent field and the cavity mode.

We first consider the scheme of atomic scattering in the cavity with
resonant frequency $\omega _{c}=\omega _{L}-\Omega (0)$ (A subsection).
Using Eqs. (18)-(22) we arrive the quantum state after the interaction 
\[
\langle x\mid \Psi (t)\rangle =\frac{f(x)}{\sqrt{2}}\sum_{n\geq
0}[(ac_{n}-e^{i(\varphi _{L}-\varphi _{c})}bc_{n+1})\exp (i\alpha (x,t)\sqrt{%
n+1}\sin kx)\mid N_{n}^{(+)}\rangle + 
\]
\begin{equation}
(ac_{n}+e^{i(\varphi _{L}-\varphi _{c})}bc_{n+1})\exp (-i\alpha (x,t)\sqrt{%
n+1}\sin kx)\mid N_{n}^{(-)}\rangle ]-f(x)bc_{0}e^{-i\varphi _{L}}\mid \Phi
_{2}\rangle \mid 0\rangle ,
\end{equation}
where we have introduced the following parameter 
\begin{equation}
\alpha (x,t)=\frac{g}{2}\int_{0}^{t}d\tau \left( 1-\frac{\delta }{\sqrt{%
\delta ^{2}+4\lambda ^{2}u_{L}^{2}(x,\upsilon _{z}\tau )}}\right)  \label{33}
\end{equation}
and the interaction time $t=L/\upsilon _{z}$ is the time the atom needs to
cross the resonator of length $L$.

We consider an atomic wave packet of width much smaller than the cavity mode
wavelength (more exactly, the width $\Delta x\ll 2\pi /k$) passing through
the node of the field. This assumption allows us to replace the mode
function by its expansion $\sin $ $(kx)\approx kx.$ Moreover, we omit also
the transfers profile dependence of the mode function of the coherent field,
that gives $\alpha =\frac{1}{2}gt(1-\delta /\Omega (0)).$ As a result the
probability $W_{1}(\overline{p})$ for funding an atom with scaled the
momentum $\overline{p}=p/\hbar k$ in units of photon momenta is obtained in
the following form 
\begin{eqnarray}
W_{1}(\overline{p}) &=&\frac{1}{2}\sum_{n\geq 0}[\mid ac_{n}-e^{i(\varphi
_{L}-\varphi _{c})}bc_{n+1}\mid ^{2}W^{0}(\overline{p}+\alpha \sqrt{n+1})+ \\
&\mid &ac_{n}+e^{i(\varphi _{L}-\varphi _{c})}bc_{n+1}\mid ^{2}W^{0}(%
\overline{p}-\alpha \sqrt{n+1})]+b^{2}\mid c_{0}\mid ^{2}W^{0}(\overline{p}),
\nonumber
\end{eqnarray}
where the probability $W^{0}(\overline{p})$ describes the initial
distribution of the atoms before they enter the light fields.

The other resonant case (B subsection), when $\omega _{c}=\omega _{L}+\Omega
(0)$ can be considered in the similar way. The corresponding probability of
atomic deflection $W_{2}(\overline{p})$ is calculated as 
\begin{eqnarray}
W_{2}(\overline{p}) &=&\frac{1}{2}\sum_{n\geq 0}[\mid ac_{n+1}-e^{i(\varphi
_{L}-\varphi _{c})}bc_{n}\mid ^{2}W^{0}(\overline{p}-\beta \sqrt{n+1})+ \\
&\mid &ac_{n+1}+e^{-i(\varphi _{L}-\varphi _{c})}bc_{n}\mid ^{2}W^{0}(%
\overline{p}+\beta \sqrt{n+1})]+a^{2}\mid c_{0}\mid ^{2}W^{0}(\overline{p}),
\nonumber
\end{eqnarray}
where the interaction parameter is equal to $\beta =\frac{1}{2}gt(1+\delta
/\Omega (0))$.

The expressions (33) and (34) demonstrate the multipeak structures of the
distributions centered at the discrete atomic momenta with a spacing of $%
\alpha \hbar k$ or $\beta \hbar k$. It is interesting that the magnitudes of
these spacing are proportional to the probabilities of finding the atomic
states $\mid g\rangle $ or $\mid e\rangle $ in the dressed state $\mid \Phi
_{1}\rangle $ which are $a^{2}$ or $b^{2}.$ For a sufficiently large
detuning $\delta \gg \lambda u_{L}(0)$ we have $\alpha \rightarrow 0$ and $%
\beta \rightarrow gt$, and also $a=1,b=0$. In consequence, the distribution
(33) contains only the central peak $W_{1}(\overline{p})=W^{0}(\overline{p})$
while $W_{2}(\overline{p})$ becomes equal to the well known result for the
momentum distribution of a deflected atom which enters the cavity in the
ground state $\mid g\rangle $ in the absence of a coherent field [1]. The
other point that should be mentioned is that the superposition of the two
atomic states contributes to $W_{1}(\overline{p})$ and $W_{2}(\overline{p})$
as the interference terms. Such effects have been studied in [12] for the
deflection of atoms that are initially prepared in a superposition state of
a single cavity mode.

It should be mentioned again that the momentum distribution $W_{1}(\overline{%
p})$ describes the atomic deflection due to the multiphoton Raman scattering
process. We comment this result for the simplest situation where the cavity
field is initially in the vacuum state: $c_{0}=1,\,c_{n}=0,\,n\neq 0,$ \
i.e. the potential is produced by the vacuum of the cavity field. In this
case the momentum distribution reads 
\begin{equation}
W_{1}(\overline{p})=\frac{a^{2}}{2}(W^{0}(\overline{p}+\alpha )+W^{0}(%
\overline{p}-\alpha ))+b^{2}W^{0}(\overline{p})
\end{equation}
and shows three peaks at $p=0$ and $p=\pm \alpha \hbar k.$ Expanding the
initial state as $\mid g\rangle =a\mid \Phi _{1}\rangle -be^{-i\varphi
_{L}}\mid \Phi _{2}\rangle $, consider the dynamics of two component. Note
that the interaction with cavity mode lead to the decay of dressed state $%
\mid \Phi _{1}\rangle \rightarrow \mid \Phi _{2}\rangle ,$ with the
amplitude proportional to $a^{2}.$ The corresponding Raman transition $\mid
g\rangle \mid o\rangle \rightarrow \mid e\rangle \mid 1\rangle $ with
radiation of $\omega _{L}-\Omega $ photon forms the peaks at $p=\pm \alpha
\hbar k$ of the distribution (35). In this way the central peak is
determined by $\mid \Phi _{2}\rangle $ state overlap of $\mid g\rangle $
with the amplitude $b^{2}$. To illustrate this result we consider a spatial\
distribution f(x), which is Gaussian with width $\triangle x.$ Its
corresponding initial momentum distribution is equal to $W^{\left( 0\right)
}(\overline{p})=k\triangle x\exp (-k^{2}\triangle x^{2}\overline{p}^{2})/%
\sqrt{\pi }$. In Fig 3 we depict the momentum distribution (35) for two
values of the interaction parameter $d=4\lambda ^{2}u_{L}{}^{2}/\delta ^{2}$
and for strong-coupling regime. In Fig.4 for comparison, we show
distribution $W_{2}(\overline{p})$ for the other resonant configuration $%
(\omega _{c}=\omega _{L}+\Omega )$ considering interaction of moving
Gaussian wavepacket with a single mode vacuum field in a cavity $%
(c_{0}=1,\,c_{n}=0,\,n\neq 0).$ Note that in Figs. 3, 4 attainable for the
experiments parameters are used. In fact, according to the previous analysis
the range of the interaction parameter $gt$ should be $\max [1,g/\lambda
]\ll gt\ll \max [g/\gamma ,g/k]$, that is mainly restricted by the ratios of
the coupling constant on the decay rates. So far, several experiments in
cavity \ QED [16-18] have investigated the interaction of an atom with
electromagnetic field in the strong coupling regime, in particular, with the
following parameters $g/\gamma =6,g/k=11$ [18].

\section{Conclusion}

In conclusion, we have presented a novel model in atomic optics based on the
doubly-dressed states. We suggest that various effects of atomic optics can
be explained from such simple model involving the dressing of the strong
field-atom dressed states by the cavity field. Really, it is shown that the
cavity with fixed number of photons creates a barrier and a well potential
for the external motion of the atom corresponding to the doubly-dressed
states $\mid N_{n}^{(\pm )}\rangle $ (in the configuration A) and $\mid
R_{n}^{(\pm )}\rangle $ (in the configuration B). Such model generalizes the
well-known approach in atomic optics in quantized light fields on the case
involved also the strong coherent field. In distinct to the standard atomic
scattering problems in which the atom-cavity mode couplings are constants,
in the presented model the corresponding couplings are the functions of the
detuning and the resonant strong-field Rabi frequency. We have applied this
model to the deflection of two-level atomic wave packets. Although the
primary application of the model was concerned with two-level atom, the
analogous results may be obtained for various atomic configurations.

{\bf Acknowledgments}

I acknowledge helpful discussions with H. K. Avetissyan, M. Freyberger and
W. P. Schleich. This work was supported by ISTC grant N A-353.

\begin{center}
{\bf References}
\end{center}

1. M. Freyberger, A. M. Herkommer, D. S. Kr\"{a}hmer, E. Mayr and W. P.
Schleich, Advances in Atom. Mol., and Opt. Phys. {\bf 41 }(1999) 443.

2. W. Walther, Phys. Scripta {\bf 76 }(1998) 138.

3. S. Haroche, Phys. Scripta {\bf 76 }(1998) 159.

4. B.-G. Englert, J. Schwinger, A. O. Barut and M. O. Scully, {\it Europhys.
Lett.}{\bf \ 14} (1991) 25

5. M. O. Scully, G. M. Meyer and H. Walther, {\it Phys.Rev.Lett. }{\bf 76}
(1996) 4144.

6. G. M. Meyer, M. O. Scully and H. Walther, {\it Phys.Rev. }{\bf A 56}
(1997) 4142.

7. \ M. Loeffler, G. M. Meyer, M. Schroeder, M. O .Scully and H. Walther, 
{\it Phys.Rev.} {\it \ }{\bf A 56} (1997) 4153.

8. M. Schroeder, K. Vogel, W. P. Schleich, M. O. Scully and H. Walther, {\it %
Phys.Rev.} {\it \ }{\bf A 56} (1997) 4164.

9. \ J. Dalibard and C. Cohen-Tannoudji,{\it \ J.Opt.Soc.Am}. {\bf B 2}
(1985) 1707.

10. P. Meystre, E. Schumacher and S. Stenholm, {\it Opt.Commun.} {\bf 73 }%
(1989) 443.

11. A. M. Herkommer, V. M. Akulin and W. P. Schleich, {\it Phys.Rev.Lett.} 
{\bf 69} (1992) 3298.

12. M. Freyberger and A. M. Herkommer, {\it Phys.Rev.Lett. }{\bf 72} (1994)
1952.

13. F. Treussart, J. Hare, L. Collot, V. Lefevre, D. S. Weiss, V.
Sandoghdar, J. M. Raimond, and S. Haroche, {\it Opt.Lett. }{\bf 9} (1994)
1651.

14. D. M. Ciltner, R. W. McGowan and S. Lee, {\it Phys.Rev.Lett}. {\bf 75}
(1995) 2638.

15. A. Lezama, Yifu Zhu, M. Kanskar and T. W. Mossberg, {\it Phys.Rev.}{\bf %
\ A 41} (1990) 1576.

16. M. Brune, F. Schmidt-Kaler, A. Maali, J. Dreyer, E. Hagley, J. M.
Raimond, and S. Haroche, {\it Phys.Rev.Lett. }{\bf 76} (1996) 1800.

17. C. J. Hood, M. S. Chapman, T. W. Lynn, and H. J. Kimble, {\it %
Phys.Rev.Lett. }{\bf 80} (1998) 4157.

18. P. M\"{u}nstermann, T. Fischer, P. Maunz, P.W. H. Pinkse, and G. Rempe, 
{\it Phys.Rev.Lett. }{\bf 82} (1999) 3791.

\begin{center}
{\bf Figure captions\smallskip }
\end{center}

\smallskip

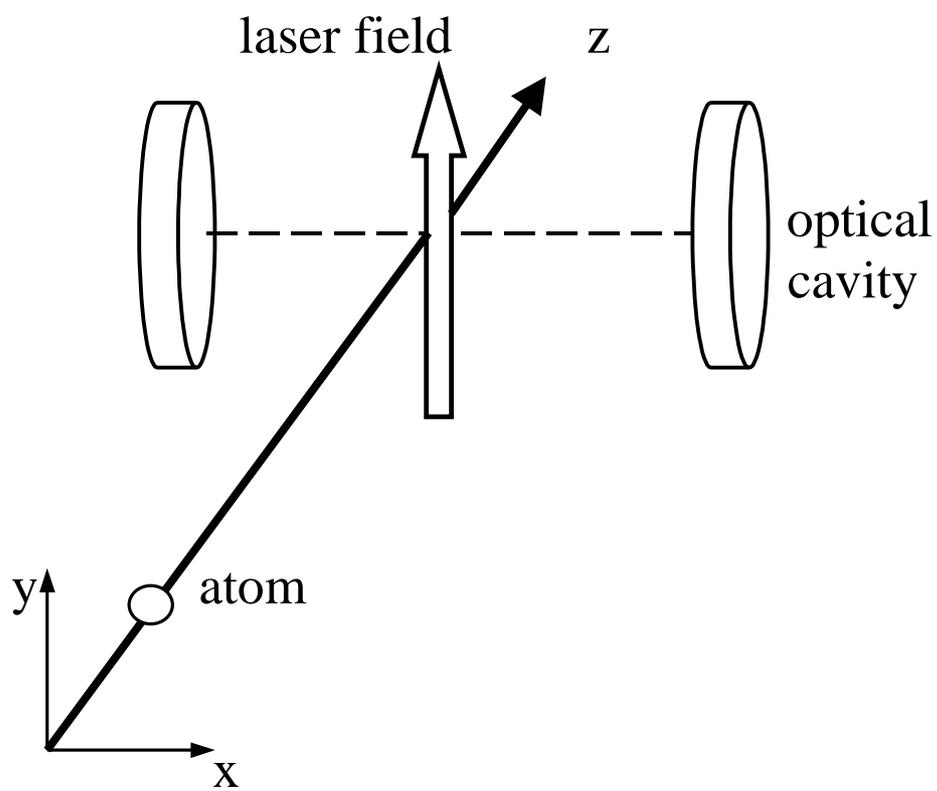
Fig.1. Sketch of the experimental setup. A two-level atom in the ground state%
$\mid g\rangle $ moving along z-direction enters a cavity with both a
single-quantized mode and a classical laser field. Pump laser beam crosses
the cavity axis.

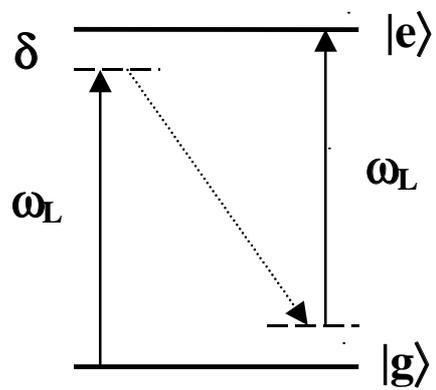
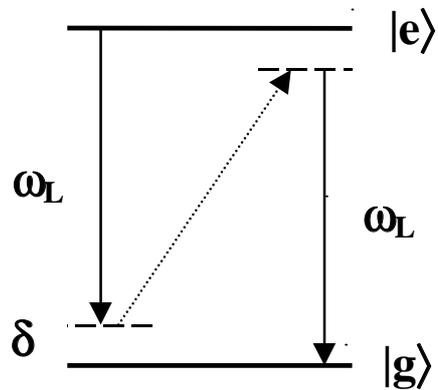
Fig. 2. Schematic diagrams for the Raman processes showing the transfer of
population between the ground and the excited states. Stimulated emission
(a) and absorption (b) of photon at frequency $\omega _{L}-\Omega (0)\simeq
2\omega _{L}-(\omega _{e}-\omega _{g})$ (dashed arrows) of the field mode
occur due to the interaction of two-level atom with laser field (straight
arrows).

Fig. 3. Momentum distribution of the atomic wavepacket after interaction
with $\omega _{c}=\omega _{L}-\Omega (0)$ mode vacuum field in the cavity.
The parameters are: $k\Delta x=1,\,gt=50,\,d=1(a),\,d=0.6(b).$

Fig. 4. Momentum distribution $W_{2}(\overline{p})$ of the atomic wavepacket
passing through the empty cavity calculated from Eq. (34) for the two
interaction parameters $d=1(a)\,$and$\,\,d=1.5(b).$The other parameters are: 
$k\Delta x=1,\,gt=50.$

\smallskip

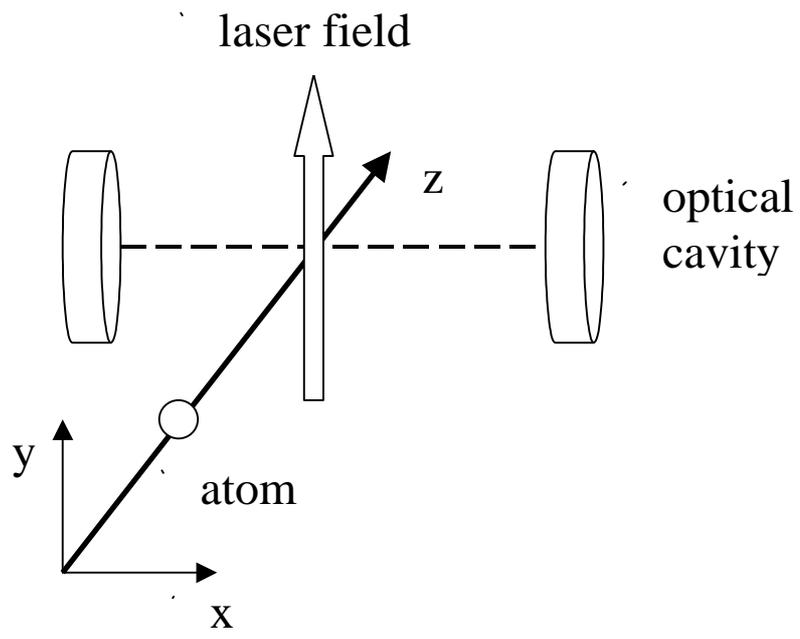
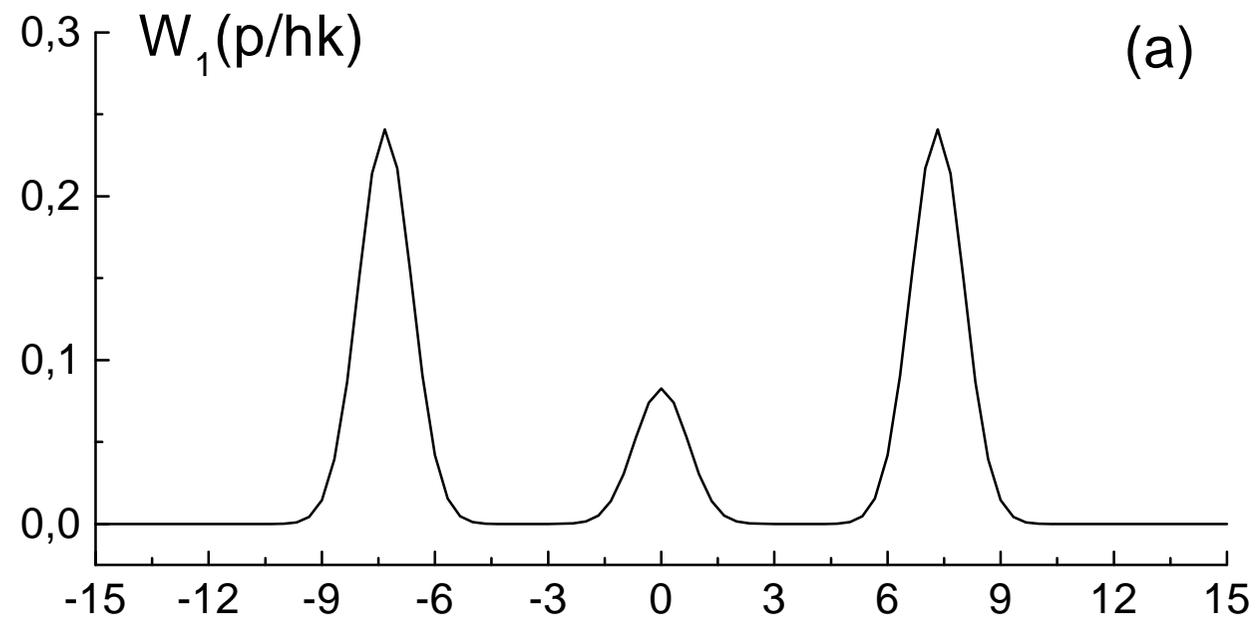
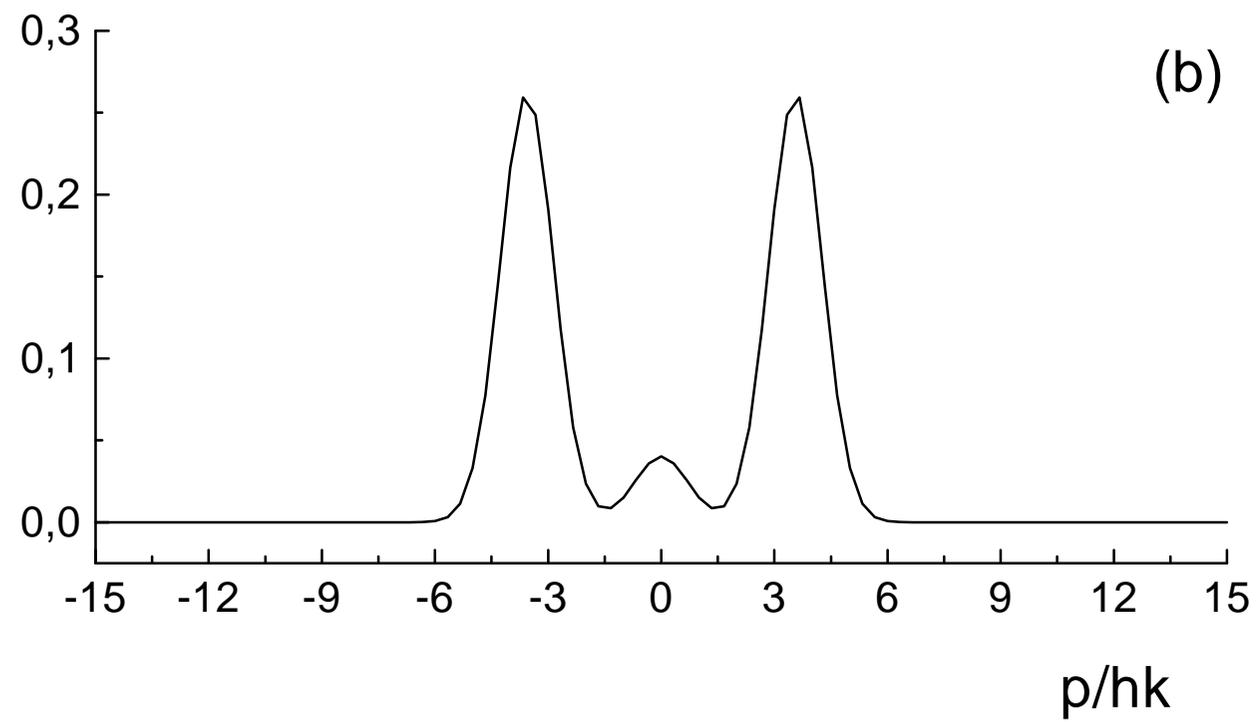
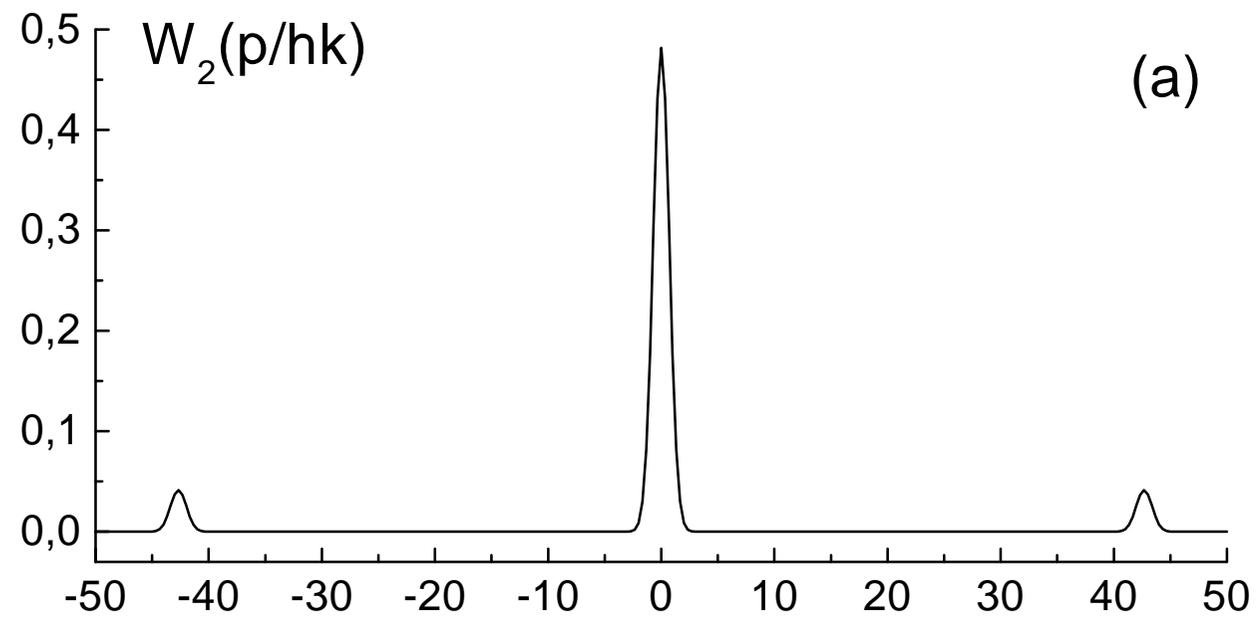
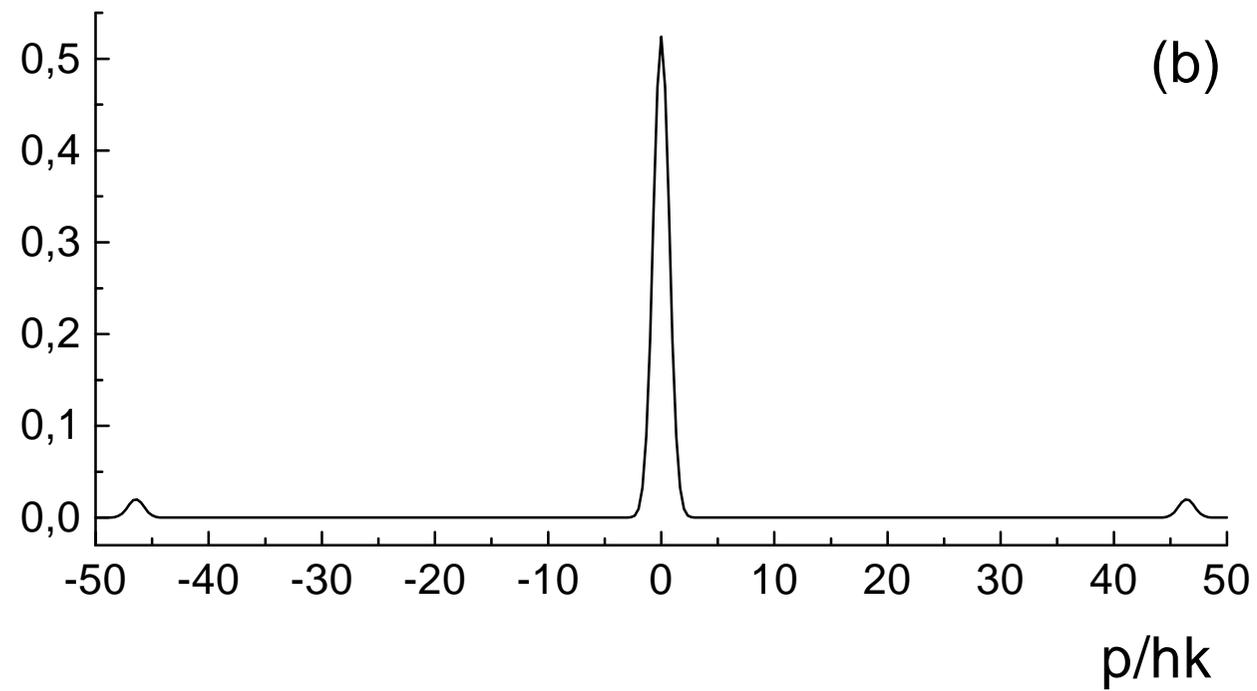
\end{document}